# Ghost Branch Photoluminescence From a Polariton Fluid Under Nonresonant Excitation

Maciej Pieczarka [1,*], Marcin Syperek [1], Łukasz Dusanowski [1], Jan Misiewicz [1], Fabian Langer [2], Alfred Forchel [2], Martin Kamp [2], Christian Schneider [2], Sven Höfling [2,3], Alexey Kavokin [4,5], Grzegorz Sęk [1]

[1] *Laboratory for Optical Spectroscopy of Nanostructures, Department of Experimental Physics, Wrocław University of Technology, Wybrzeże Wyspiańskiego 27, 50-370 Wrocław, Poland*
[2] *Technische Physik, University of Würzburg and Wilhelm-Conrad-Röntgen-Research Center for Complex Material Systems (RCCM), Am Hubland, D-97074 Würzburg, Germany*
[3] *SUPA, School of Physics and Astronomy, University of St. Andrews, St. Andrews KY16 9SS, United Kingdom*
[4] *Spin Optics Laboratory, Saint Petersburg State University, 1, Ulianovskaya, 198504, St Petersburg, Russia*
[5] *Physics and Astronomy School, University of Southampton, Highfield, Southampton SO17 1BJ, UK.*
[*]corresponding author: maciej.pieczarka@pwr.edu.pl

**Abstract**

An expanding polariton condensate is investigated under pulsed nonresonant excitation with a small laser pump spot. Far above the condensation threshold we observe a pronounced increase in the dispersion curvature with a subsequent linearization of the spectrum and strong luminescence from a ghost branch orthogonally polarized with respect to the linearly polarized condensate emission. Polarization of both branches is understood in terms of spin-dependent polariton-polariton scattering. The presence of the ghost branch has been confirmed in time-resolved measurements. The effects of disorder and dissipation in the photoluminescence of polariton condensates and their excitations are discussed.

Exciton polaritons are composite bosons consisting of strongly coupled microcavity photons and quantum well excitons. They are able to form a novel class of condensates (for recent reviews see, e.g. [1], [2]). Despite their nonequilibrium and dissipative nature, they behave as condensates of weakly interacting bosons. Collective phenomena like condensation [3], off diagonal long range order [4], [5], topological excitations [6] or superfluid features in propagation of polariton flows [7] have been demonstrated in such systems. A spectacular consequence of the parametric scattering in polariton gases is appearance of non-parabolic scattering bands, including normal branches (NBs) and ghost branches (GBs), the latter ones populated by the virtual off-branch exciton-polaritons [7,8]. Excitations of polariton condensates are expected to be characterized by linearized dispersions of the Bogoliubov-like spectra [9]. The fluid excitations at low momenta are expected to behave as collective sound waves rather than single particles. Recently, the linear dispersion NB and the GB of a resonantly pumped polariton condensate [10] have been observed in a four-wave mixing experiment.

The experiment providing a direct access to the dispersion of excitations of spontaneously formed condensates in polariton systems is photoluminescence (PL) under nonresonant excitation. The nonresonant pumping can be realized either with a detuned laser or with the current injection [11,12]. In the nonresonant pumping scheme, incoherent excitons are generated and relax, subsequently feeding the reservoir and governing the dynamics of the system [13]. The relaxation of created hot excitons involves multiple scattering processes, which destroy the coherence and phase of the excitation, ensuring that these properties are not inherited by the condensate, in contrast to the resonant excitation case. At the same time, the incoherent reservoir causes additional decoherence [14] forming a repulsive potential [15], which shapes the condensate spatially and spectrally. Moreover, it affects significantly the excitation spectrum of the condensate [16]. Bogoliubov dispersions have been reported in this excitation scheme, however, no fingerprints of the GB have been observed yet [17], [18]. This fueled the debate on the possibility of observing the GB in nonresonant PL [19] where the strong emission of the condensate occurs and that one being related to the NB can easily mask the GB signal. Furthermore the incoherent reservoir might affect the Bogliubov-like spectra of the excitations of the condensate. The resonant experiment done by V. Kohnle et. al. in [10] is in stark contrast to nonresonant pumping scheme, since the polariton condensate formation occurred in absence of incoherent reservoir and the collective excitations were created by a resonant trigger pulse.

In this Letter, we report on the direct observation of a PL signal of the GB together with a pronounced linearization of the dispersion of an expanding polariton condensate under nonresonant excitation. We verify the observation via polarization- and time-resolved experiments, which elucidate the origin of the observed features.

Details on the experimental setup and the used sample are given in the supplementary material [20].

In our experiment the pulsed pumping laser was focused to a diffraction limited Gaussian shape of about 2 μm in diameter. The PL intensity as a function of the excitation power for two linear and perpendicular polarization detections, co-linear and cross-linear to the polarization axis of the polariton condensate, is shown in Fig. 1(a). The data is extracted from time-integrated data, hence the values are averaged over many excitation pulses. At low excitation densities, the lower polariton branch (LP) dispersion is formed as can be seen in Fig 1(b). With a further increase of the excitation power, we observe a distinct nonlinear threshold in the intensity dependence on pumping shown in Fig. 1(a), consistent with the formation of a flat dispersion of a polariton condensate [Fig. 1(c)] at the threshold. This kind of dispersion is predicted [16] when exciting a polariton condensate by a small excitation spot. Most likely, it is caused by the localization of polaritons by the disorder potential in our case. The condensation threshold is also manifested by a decrease of the linewidth of the measured emission which drops down by a factor of 5 at threshold, and by a blueshift of the emission peak of about 3 meV [20]. The energy shift of the propagating condensate created nonresonantly is composed of the kinetic energy of polaritons, repulsive interactions within the condensate and a term coming from the interaction of polaritons with pump-spot-induced reservoir of noncondensed quasi-particles [15,21]. Hence, the condensate energy is shifted above the minimum of the bare cavity mode, while the strong coupling is preserved.

The microcavity polaritons subject to a natural disorder potential which originates from the width and composition fluctuations of the quantum wells and Bragg mirrors [22]. We observe polariton trapping by the disorder potential as one can see in Figure 2. Localization of polaritons in real space has been observed at moderate pump densities in our experiment. Bright spots of the luminescence located outside the pump area are visible in Fig. 2(a), in agreement with the flat parts of the polariton dispersion observed in the reciprocal space [Fig. 2(c)] [23]. The most striking phenomena have occurred at higher polariton densities. First, we observed an evident spreading of the polariton cloud to distances much larger than the pump spot – Fig. 2(b). This can be interpreted in terms of ballistic flow of the polariton condensate pumped above the percolation threshold in a 2D disorder potential. Furthermore, a pronounced linearization of the polariton emission and a clear evidence of the GB emission at the negative energies (below the condensate emission energy) have been observed, as presented in Fig. 2(d). The GB becomes visible at pump densities around $20P_{th}$ at different detunings and at different spot positions at the sample surface. The buildup of the GB luminescence could be attributed to the off-branch multiple-scattering processes in the polariton gas [7]. However, the intensity of GB is strong and comparable to the NB and we detected a change in the dispersion curvature, indicating a collective phenomenon of mixing the NB and GB [10]. The linear spectrum and the appearance of the GB are fingerprints of collective Bogoliubov-like excitations of a propagating polariton fluid. Similar features can be induced resonantly in a propagating condensate in the supersonic regime in the presence of obstacles which disturb the flow [24], [25]. Here, one can observe interference ripples in the real space image [Fig. 2(d)], which is a signature of nonradial propagation of scattered polariton waves [26]. In our sample the magnitude of the disorder potential fluctuations can be as high as 3 meV, being comparable to the observed blueshift, hence it is believed to be the source of excitations in expanding polariton fluid, behaving as a static scattering landscape. The PL signal coming from ghost excitations is distributed on macroscopic distances, outside of the pump spot [20]. Additionally, one has to note a slight anisotropy of the emission intensity distribution over the dispersion branches, which reflects a non-uniform disorder potential in the real space.

Let us now discuss the polarization characteristics of emitted photons. One important property of polaritons is their spin fine structure inherited from excitons and photons. The optically active LPs have the spin projection $S=\pm 1$ on the growth axis direction. Due to the anisotropic spin interactions the minimization of the system energy favors the superposition of spin up and spin down polaritons at the condensation threshold, resulting in a linear polarization buildup [27]. In our experiment the nonlinear threshold is accompanied by an abrupt increase of the degree of linear polarization (DOLP) around $\mathbf{k}=0$, presented in Fig. 3(a) (where $DOLP = I_\parallel - I_\perp / I_\parallel + I_\perp$; $I_{\parallel,\perp}$ are the intensities detected in two orthogonal linear polarizations). The condensate forms at each point in the real space with the same direction of linear polarization vector. Our sample is disordered, which is why the polarization vector of the condensate is always pinned to one of the crystal axes [28,29]. For higher pumping levels we observe depinning of the polarization vector, manifested in a decrease of the total DOLP with respect to the pump power [30]. This data is extracted from the time-integrated measurements, so the measured DOLP values are averaged over hundreds of time evolutions, resulting in the lower value of the DOLP compared to each individual realization [31].–Taking into account the polarization properties of the excitations, there are NB and GB states available for polaritons with collinear and orthogonal to the condensate polarization vector. However, considering the occupation of the NB and GB in our experiment, one should expect the NB to be polarized as the condensate and the GB signal to be orthogonal to it. This is due to the physical origin of the two coupled branches of excitations. While the NB is populated due to the non-zero effective temperature of the polariton gas, the GB is populated due to the depletion of the condensate by polariton-polariton scattering similarly to the off-branch scattered states observed at resonant pumping in [7]. According to the polarization selection rules [31], [32], scattering of linearly polarized polaritons produces polaritons having an orthogonal linear polarization, preferentially, as a result of spin-dependent polariton-polariton scattering [20]. Indeed, our polarization-resolved measurements have shown the GB to be orthogonally polarized with respect to the condensate polarization axis. The results are shown in Figures 3(b), where signal from separate GB and NB as a function of polarization angle is shown, and 3(e), where we plot a DOLP map created from a direct overlap of the two dispersions from Fig. 3(c) and (d) and calculating the polarization degree for each pixel of the recorded data. A large area of a positive DOLP is observed for the NB, which comes from the background created by the full time evolution of the signal after the pump pulse. More importantly, very distinct inversed polarization of the GB is clearly visible in the DOLP map. It is worth noting that while the optical nonlinearities of the cavity itself can generate an analogue of complex quantum fluid phenomena in the paraxial geometry [33], however the large value of DOLP and linear polarization inversion for GB is a direct evidence of the polariton-polariton scattering responsible for the population of GB. Clearly, this cannot be an effect of the linear polarization splitting in the cavity, which could rather give a rise to the buildup of the circular polarization of emission [34].

All the data discussed above has been obtained in the time-integrated measurements. The main drawback of this kind of approach in experiments with pulsed excitation is a loss of the information about the complex dynamics of the system after each excitation pulse, in particular, it leads to the time-averaging of the blueshift of the polariton dispersion branch. This has been addressed in Ref. [10], where the authors have observed an overestimated speed of sound in the time-integrated picture as well as an asymmetry between slopes of the NB and GB. One can observe a similar asymmetry in our experiment, where the NB has a greater slope than the GB - see Figs. 3(c-d). In order to rule out the artifacts in our dispersions which might arise due to the time integration, we have performed the time-resolved scanning of the measured dispersion of a propagating polariton condensate at the detuning of –2 meV, and we studied in detail the dynamics of the polariton luminescence after the pump pulse.

Approximately 50 ps after the pump pulse arrival we have observed a simultaneous and distinct appearance of the positive and negative branches in the photoluminescence. We exclude here the apparatus effect, as the collected signal lasts much longer than the response resolution of the setup. A snapshot at 66 ps after the pulse arrival is shown in Fig. 4(a). One can notice that the NB and GB have now comparable slopes in this time-resolved picture. However, the extracted propagation velocity is still somewhat larger than the speed of sound calculated from the condensate blueshift from Fig. 4(a): the velocity taken from the slope of the time-resolved experiment is $v_{slope} \approx 1.95$ µm/ps and is greater than the velocity of $v_{blue} \approx 1.45$ µm/ps calculated from the blueshift at $\boldsymbol{k} = \boldsymbol{0}$, according to definition $c = \sqrt{\frac{U}{m_{pol}}}$, where $U$ is the blueshift and $m_{pol}$ is the LP effective mass.

The $v_{slope}$ corresponds to the polariton mean field energy of $U \approx 3.8\ meV$, being four times larger than the observed temporal blueshift. It has to be noted, that the used excitation scheme creates a condensate with a nontrivial spatial distribution of wave vectors [15]. Our case is far from the steady state of a static condensate and in this experimental configuration the pump pulse creates moving condensate, characterized by the outward coherent flow and disorder scattering of the polaritonic waves. This is why the condensate at zero wave vector is more likely to be created via scattering on multiple defect centers [35] resulting in lower blueshift than for a static condensate. We can also speculate on the spatially varying local Doppler shift of created excitations which might be responsible for the observed distinct slopes, adding additionally the local condensate velocity to the slope of the dispersion.

The observed emission of the condensate and the GB are present at the energy very close or even above the energy of the cavity mode in our sample, giving rise to the question if these phenomena occur still in the strong coupling regime. In the time-resolved spectrum right after the pump pulse arrival, we observe a strong emission corresponding to the bare cavity mode at the time scale below the temporal resolution of our setup as shown in Fig. 4(b). All the recorded dispersion features including the linearized NB and the orthogonally polarized GB signal occur later in time, where the weak coupling lasing has completely vanished. We conclude that these spectral features are characteristics of a polariton condensate formed in the strong coupling regime. In fact, in the time-resolved measurements we observe a transition from weak to strong coupling similar to the one reported in [36]. Furthermore, the signal related to GB and the pronounced slope of the NB is clearly distinguishable from possible weakly coupled lasing from localized states, which may appear in their energetic proximity. The large distance spread of the emission cloud from the pump spot is achievable only for low mass particles (weak coupling lasing spot radius should evolve differently with pumping power [20], [37]) and the depinning of the polarization vector of the condensate NB is exclusive for interacting particles being a manifestation of strong coupling preservation [30].

We believe that we were able to observe very clearly the renormalization of the NB and the buildup of the GB partly due to the relatively modest quality factor of our sample, around 1000 [20], which was low enough to enable efficient extraction of photons outside the cavity and high enough to preserve the conditions for the polariton condensate formation. Moreover, the intrinsic large disorder of the cavity enhanced the elastic scattering favorable for the condensate excitations formation.

In conclusion, we report on the first observation of a polariton ghost branch in a photoluminescence experiment under nonresonant excitation. Even though the system in our approach was far from equilibrium and from the steady state regime, we have observed a distinct renormalization of the polariton dispersion characterized by the Bogoliubov-like normal and ghost branches. The origin of the observed dispersion branches has been identified based on the observed orthogonal linear polarizations of the branches and on the time-resolved dispersions showing temporal symmetric slopes for the normal and ghost branches of the polariton condensate excitations.

**Acknowledgments**

We would like to thank Iacopo Carusotto for critical reading of this manuscript and for valuable discussions. The authors acknowledge the financial support from the bilateral project of Deutsche Forschungsgemeinschaft (project named LIEPOLATE) and Polish Ministry of Science and Higher Education (project No. DPN/N99/DFG/2010). The experiment was partially performed within the NLTK infrastructure, Project No. POIG. 02.02.00-003/08-00. AK acknowledges the financial support from the Russian Ministry of Education and Science (Contract No.11.G34.31.0067).

**References**

[1]   T. Byrnes, N. Y. Kim, and Y. Yamamoto, Nat Phys **10**, 803 (2014).

[2]   I. Carusotto and C. Ciuti, Rev. Mod. Phys. **85**, 299 (2013).

[3]   J. Kasprzak, M. Richard, S. Kundermann, A. Baas, P. Jeambrun, J. M. J. Keeling, F. M. Marchetti, M. H. Szymańska, R. André, J. L. Staehli, V. Savona, P. B. Littlewood, B. Deveaud, and L. S. Dang, Nature **443**, 409 (2006).

[4]   J. Fischer, I. G. Savenko, M. D. Fraser, S. Holzinger, S. Brodbeck, M. Kamp, I. A. Shelykh, C. Schneider, and S. Höfling, Phys. Rev. Lett. **113**, 203902 (2014).

[5]   F. Manni, K. G. Lagoudakis, R. André, M. Wouters, and B. Deveaud, Phys. Rev. Lett. **109**, 150409 (2012).

[6]   G. Nardin, G. Grosso, Y. Léger, B. Piętka, F. Morier-Genoud, and B. Deveaud-Plédran, Nat. Phys. **7**, 635 (2011).

[7]   P. Savvidis, C. Ciuti, J. Baumberg, D. Whittaker, M. Skolnick, and J. Roberts, Phys. Rev. B **64**, 075311 (2001).

[8]   J. Zajac and W. Langbein, arXiv Prepr. arXiv1210.1455 (2012).

[9]   L. Pitaevskii and Stringari S., *Bose-Einstein Condensation* (Oxford University Press, New York, 2003).

[10]  V. Kohnle, Y. Léger, M. Wouters, M. Richard, M. T. Portella-Oberli, and B. Deveaud-Plédran, Phys. Rev. Lett. **106**, 255302 (2011).

[11]  C. Schneider, A. Rahimi-Iman, N. Y. Kim, J. Fischer, I. G. Savenko, M. Amthor, M. Lermer, A. Wolf, L. Worschech, V. D. Kulakovskii, I. A. Shelykh, M. Kamp, S. Reitzenstein, A. Forchel, Y. Yamamoto, and S. Höfling, Nature **497**, 348 (2013).

[12]  P. Bhattacharya, T. Frost, S. Deshpande, M. Z. Baten, A. Hazari, and A. Das, Phys. Rev. Lett. **112**, (2014).

[13] F. Tassone, C. Piermarocchi, and V. Savona, Phys. Rev. B **56**, 7554 (1997).

[14] J. Schmutzler, T. Kazimierczuk, Ö. Bayraktar, M. Aßmann, M. Bayer, S. Brodbeck, M. Kamp, C. Schneider, and S. Höfling, Phys. Rev. B **89**, 115119 (2014).

[15] M. Wouters, I. Carusotto, and C. Ciuti, Phys. Rev. B **77**, 115340 (2008).

[16] M. Wouters and I. Carusotto, Phys. Rev. Lett. **99**, 140402 (2007).

[17] S. Utsunomiya, L. Tian, G. Roumpos, C. W. Lai, N. Kumada, T. Fujisawa, M. Kuwata-Gonokami, A. Löffler, S. Höfling, A. Forchel, and Y. Yamamoto, Nat. Phys. **4**, 700 (2008).

[18] M. Assmann, J.-S. Tempel, F. Veit, M. Bayer, A. Rahimi-Iman, A. Löffler, S. Höfling, S. Reitzenstein, L. Worschech, and A. Forchel, Proc. Natl. Acad. Sci. U. S. A. **108**, 1804 (2011).

[19] T. Byrnes, T. Horikiri, N. Ishida, M. Fraser, and Y. Yamamoto, Phys. Rev. B **85**, (2012).

[20] See Supplementary Material, including Refs. [26], [31,32],[38-46].

[21] P. Cilibrizzi, A. Askitopoulos, M. Silva, F. Bastiman, E. Clarke, J. M. Zajac, W. Langbein, and P. G. Lagoudakis, Appl. Phys. Lett. **105**, 191118 (2014).

[22] V. Savona, J. Phys. Condens. Matter **19**, 295208 (2007).

[23] D. Sanvitto, A. Amo, L. Viña, R. André, D. Solnyshkov, and G. Malpuech, Phys. Rev. B **80**, 045301 (2009).

[24] I. Carusotto and C. Ciuti, Phys. Rev. Lett. **93**, 166401 (2004).

[25] A. Amo, J. Lefrère, S. Pigeon, C. Adrados, C. Ciuti, I. Carusotto, R. Houdré, E. Giacobino, and A. Bramati, Nat. Phys. **5**, 805 (2009).

[26] G. Christmann, G. Tosi, N. G. Berloff, P. Tsotsis, P. S. Eldridge, Z. Hatzopoulos, P. G. Savvidis, and J. J. Baumberg, Phys. Rev. B **85**, 235303 (2012).

[27] F. Laussy, I. Shelykh, G. Malpuech, and A. Kavokin, Phys. Rev. B **73**, 035315 (2006).

[28] J. Kasprzak, R. André, L. Dang, I. A. Shelykh, A. V. Kavokin, Y. Rubo, K. Kavokin, and G. Malpuech, Phys. Rev. B **75**, 045326 (2007).

[29] Ł. Kłopotowski, M. D. Martín, A. Amo, L. Viña, I. A. Shelykh, M. M. Glazov, G. Malpuech, A. V. Kavokin, and R. André, Solid State Commun. **139**, 511 (2006).

[30] J. Levrat, R. Butté, T. Christian, M. Glauser, E. Feltin, J.-F. Carlin, N. Grandjean, D. Read, A. V. Kavokin, and Y. G. Rubo, Phys. Rev. Lett. **104**, 166402 (2010).

[31] T. Ostatnický and A. V. Kavokin, Superlattices Microstruct. **47**, 39 (2010).

[32] T. Ostatnický, I. A. Shelykh, and A. V. Kavokin, Phys. Rev. B **81**, 125319 (2010).

[33] J. Scheuer and M. Orenstein, Science **285**, 230 (1999).

[34] P. Cilibrizzi, H. Ohadi, T. Ostatnicky, A. Askitopolous, W. Langbein, and P. Lagoudakis, Phys. Rev. Lett. **113**, 103901 (2014).

[35] M. Wouters and I. Carusotto, Phys. Rev. Lett. **105**, 020602 (2010).

[36] E. Kammann, H. Ohadi, M. Maragkou, A. V. Kavokin, and P. G. Lagoudakis, New J. Phys. **14**, 105003 (2012).

[37] H. Deng, G. Weihs, D. Snoke, J. Bloch, and Y. Yamamoto, Proc. Natl. Acad. Sci. U. S. A. **100**, 15318 (2003).

[38] G. Roumpos, M. D. Fraser, A. Löffler, S. Höfling, A. Forchel, and Y. Yamamoto, Nat. Phys. **7**, 129 (2011).

[39] J. M. Zajac, E. Clarke, and W. Langbein, Appl. Phys. Lett. **101**, 041114 (2012).

[40] A. Baas, K. G. Lagoudakis, M. Richard, R. André, L. S. Dang, and B. Deveaud-Plédran, Phys. Rev. Lett. **100**, 170401 (2008).

[41] E. Wertz, L. Ferrier, D. Solnyshkov, R. Johne, D. Sanvitto, A. Lemaître, I. Sagnes, R. Grousson, A. V. Kavokin, P. Senellart, G. Malpuech, and J. Bloch, Nat. Phys. **6**, 19 (2010).

[42] I. Shelykh, Y. Rubo, G. Malpuech, D. Solnyshkov, and A. V. Kavokin, Phys. Rev. Lett. **97**, 066402 (2006).

[43] D. Read, T. C. H. Liew, Y. G. Rubo, and A. V. Kavokin, Phys. Rev. B **80**, 195309 (2009).

[44] S. Schumacher, N. Kwong, and R. Binder, Phys. Rev. B **76**, 245324 (2007).

[45] P. Renucci, T. Amand, X. Marie, P. Senellart, J. Bloch, B. Sermage, and K. V. Kavokin, Phys. Rev. B **72**, 075317 (2005).

[46] M. Romanelli, C. Leyder, J. P. Karr, E. Giacobino, and A. Bramati, Phys. Rev. Lett. **98**, 106401 (2007).

**Figure captions**

Fig. 1. (a) Power dependent intensity curve of two linear polarizations, co-linear to the condensate axis (blue circles) and cross-linear to the condensate (orange circles). (b) Time integrated dispersions below (1 mW = 0.1 $P_{th}$) the threshold and (c) slightly above the threshold (5 mW = 1.66 $P_{th}$). LP branch (white line), blueshifted (green line) and bare cavity mode (red line) dispersions are shown. The color scale is linear. The measurements have been done at the cavity-exciton detuning of -0.5 meV.

Fig. 2. (a) Real space dispersion of localized polaritons at moderate densities. (b) Real space distribution of extended polariton cloud. (c) Corresponding momentum space picture to (a). (d) Onset of linear dispersion at high pumping conditions. Line description is the same as in Fig. 1. The pump spot location is indicated as a pink circle in (a) and (b). The color scale is linear.

Fig. 3. (a) DOLP as a function of the pumping power. (b) Polarizations of NB and GB branches extracted from the PL map measurements. The values are normalized. (c) Co-linear dispersion and (d) cross-linear with respect to the condensate polarization axis and (e) corresponding DOLP map of dispersion at the detuning of -0.5 meV. The DOLP scale is in the inset of (e). The intensity color scale is linear in (c) and (d).

Fig. 4. (a) Dispersion snapshot at 66 ps of the time-resolved evolution of the polariton dispersion at the detuning of -2 meV. Cavity mode and LP are shown by gray and white solid lines, respectively. The violet dotted lines are guides to the eye to highlight the similar slopes of the linear dispersion part. (b) Snapshot at the ultrashort times after the pulse arrival, presenting photon lasing in the weak coupling regime. The color scales are logarithmic.

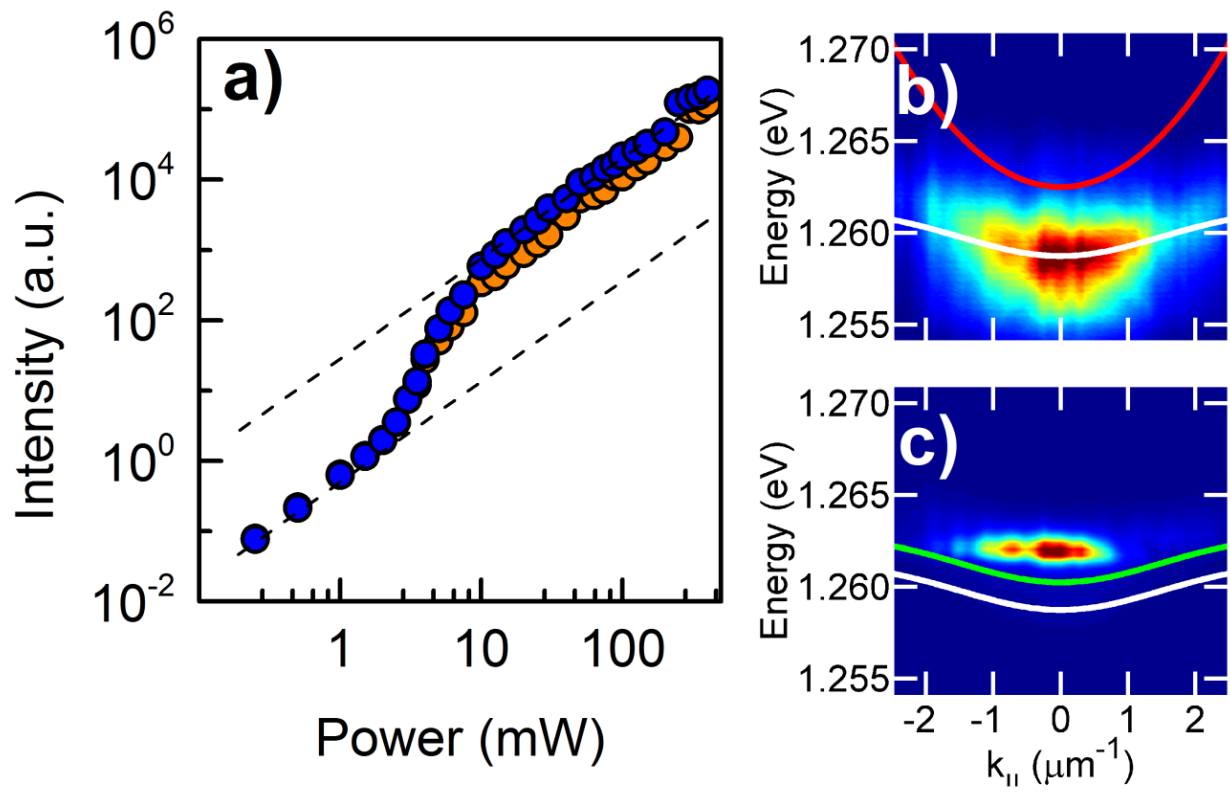

Fig. 1

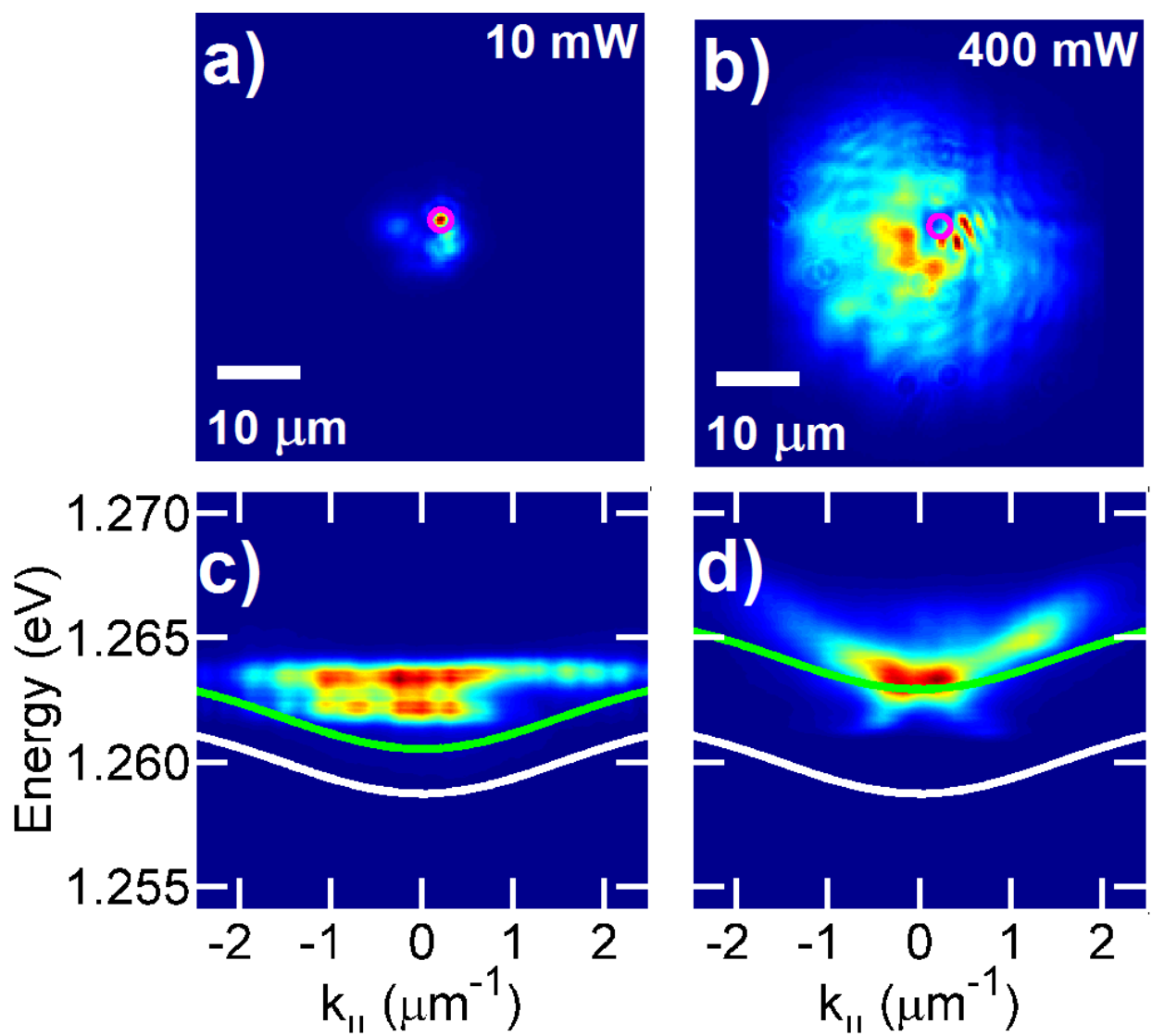

Fig. 2

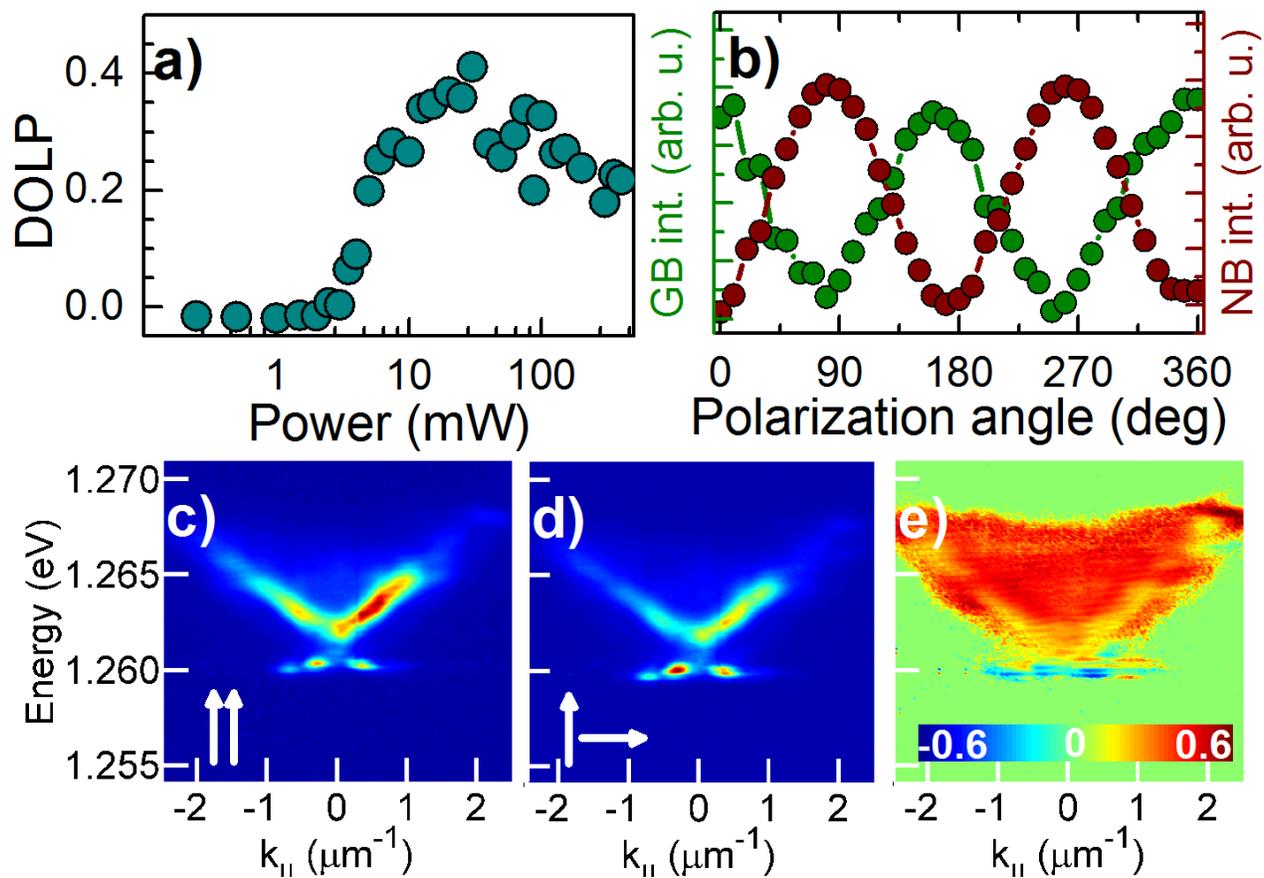

Fig. 3

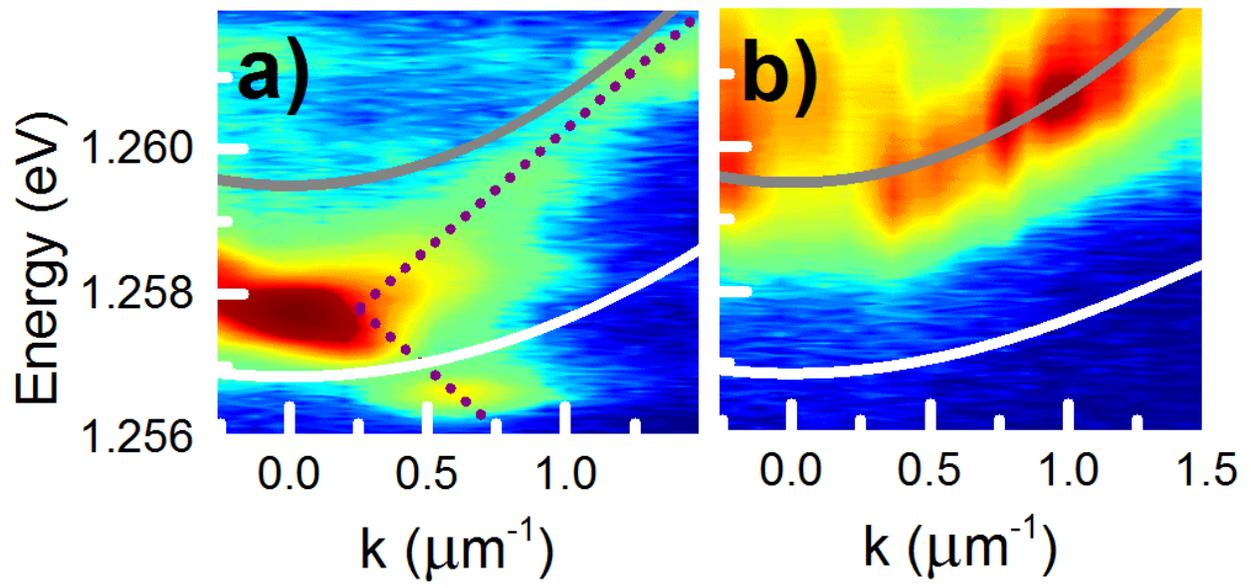

Fig. 4